\documentclass[preprint,preprintnumbers,amsmath,amssymb, superscriptaddress]{revtex4}

\usepackage{textcomp}
\usepackage{makeidx}
\usepackage{amsmath}
\usepackage{subfigure}
\usepackage{amssymb}
\usepackage{hyperref}
\usepackage{graphicx}
\usepackage[utf8]{inputenc}
\usepackage{float}
\usepackage{color}

\begin{document}

\title{The anisotropic coupling of gravity and electromagnetism in Ho\v{r}ava-Lifshitz theory}

\author{Jorge Bellorín}
\email{jbellori@gmail.com}
\affiliation{Departamento de F\'isica, Facultad de ciencias b\'asicas, Universidad de Antofagasta, Casilla 170, Antofagasta, Chile.}
\author{Alvaro Restuccia}
\email{alvaro.restuccia@uantof.cl}
\affiliation{Departamento de F\'isica, Facultad de ciencias b\'asicas, Universidad de Antofagasta, Casilla 170, Antofagasta, Chile.}
\author{Francisco Tello-Ortiz}
\email{francisco.tello@ua.cl}
\affiliation{Departamento de F\'isica, Facultad de ciencias b\'asicas, Universidad de Antofagasta, Casilla 170, Antofagasta, Chile.}


\begin{abstract}
  We analyze the electromagnetic-gravity interaction in a pure Ho\v{r}ava-Lifshitz framework. To do so we formulate the Ho\v{r}ava-Lifshitz gravity in $4+1$ dimensions and perform a Kaluza-Klein reduction to $3+1$ dimensions. We use this reduction as a mathematical procedure to obtain the $3+1$ coupled theory, which at the end is considered as a fundamental, self-consistent, theory. The critical value of the dimensionless coupling constant in the kinetic term of the action is $\lambda=1/4$. It is the kinetic conformal point for the non-relativistic electromagnetic-gravity interaction. In distinction, the corresponding kinetic conformal value for pure Ho\v{r}ava-Lifshitz gravity in $3+1$ dimensions is $\lambda=1/3$. We analyze the geometrical structure of the critical and noncritical cases, they correspond to different theories. The physical degrees of freedom propagated by the noncritical theory are the transverse traceless graviton, the transverse gauge vector and two scalar fields. In the critical theory one of the scalars is absent, only the dilaton scalar field is present. The gravity and vector excitations propagate with the same speed, which at low energy can be taken to be the speed of light. The field equations for the gauge vector in the non-relativistic theory have exactly the same form as the relativistic electromagnetic field equations arising from the Kaluza-Klein reduction of General Relativity, and are equal to them for a particular value of one of the coupling constants. The potential in the Hamiltonian is a polynomial of finite degree in the gauge vector and its covariant derivatives. 
\end{abstract}


\maketitle

\section{Introduction}

Ho\v{r}ava-Lifshitz gravity \cite{horava} is a candidate for a perturbatively renormalizable theory of quantum gravity  \cite{blas1,charmousis,visser,papazoglou,orlando,shu,benedetti,contillo,odorico,odorico1,herrero,bello2,Wang:2017brl,Shin:2017ott,pospelov}. The main idea is to break the Lorentz symmetry in order to allow higher order spatial derivative terms in the potential which improve the quantum behaviour of the theory without the introduction of ghost fields \cite{stelle}. There is no space-time metric in the formulation since there is an anisotropic scaling of the time and space coordinates. The geometrical framework is a foliation with Riemannian leaves parametrized by the time variable. The theory is manifestly invariant under the diffeomorphisms that preserve the foliation. Only reparametrizations of the time variable are allowed. Besides this Riemannian geometry the theory introduces the lapse function and shift vector as fields defined on the leaves parametrized by the time coordinate. This formulation allows the lapse function to have different dimensions than in a space-time formulation. Moreover, the overall coupling constant of the action becomes dimensionless. The Ho\v{r}ava-Lifshitz gravity may also arise from the gauging of the Newton-Cartan geometry \cite{NC}. 

In this work we consider the non-projectable version of the Ho\v{r}ava-Lifshitz gravity in $4+1$ dimensions, and perform a Kaluza-Klein (KK hereinafter) reduction to $3+1$ dimensions. The main goal will be the analysis of the gravity-vector interaction. Our final objective is to regard the resulting theory in 3+1 dimensions as a fundamental, self-consistent, theory of gravity with couplings. Hence there is no need of embedding it into a more fundamental theory. In the same way that the pure Ho\v{r}ava theory is power-counting renormalizable, we expect that the resulting coupled theory obtained by means of the dimensional reduction is also a power-counting renormalizable in $3+1$ dimensions. In this approach the fourth spatial dimension is not physical. Therefore, the dimensional reduction we employ is a mathematical approach to obtain the final and self-consistent theory in 3+1 dimension. Once the 3+1 theory with couplings has been obtained one can study any aspect of its dymanics. We remark that, although the purely gravitational nonprojectable Ho\v{r}ava theory in 3+1 dimensions is well established, the issue of its couplings to fields and other matter sources is still an open question. Our approach gives an answer in this sense. With regards to vector fields, the Ho\v{r}ava theory is frequently coupled to the relativistic energy-momentum tensor of the electromagnetic field. However, this approach is not natural in the framework of the FDiff symmetry of the Ho\v{r}ava theory. For this reason in this work we take the point of view of starting from a FDiff-invariant theory from the very beggining.

The Ho\v{r}ava theory in $4+1$ dimensions includes up to $z=4$ interaction terms in the potential. In order to preserve the invariance under diffeomorphisms on the leaves of the foliation these terms in the potential have to be constructed in terms of the $4$-dimensional Riemannian tensor, the lapse function and covariant derivatives of them \cite{blas1}. Once this requirement is satisfied one may dimensionally reduce to $3+1$ dimensions \emph{a la} KK. The same approach for GR in $4+1$ dimensions gives rise to the relativistic coupling of gravity and electromagnetism in $3+1$ dimensions. Hence our approach will determine how far the anisotropic coupling to the vector field in Ho\v{r}ava-Lifshitz theory is from the relativistic one. This is a relevant point since electrodynamics is a very well established theory. The resulting coupled Ho\v{r}ava theory in $3+1$ dimensions has both $z=1$ and $z=2,3,4$ terms. The former are the dominant ones in the physics of the large distances (infrared limit). Since we want to contrast with the standard coupling of electromagnetism with general relativity, at the level of the $3+1$ theory we take the infrared limit by discarding all the $z=2,3,4$. In this way we will end up with a theory of second order in time and spatial derivatives. We notice that for this limit we do not need the explicit expression of the $z=2,3,4$ terms of the $4+1$ theory.

The Ho\v{r}ava theory has a critical point for the dimensionless coupling constant of the kinetic term, $\lambda$, which is equal to the inverse of the spatial dimensions. This point was called the kinetic-conformal point in Ref.~\cite{bello2}. In the 4+1 theory this point correspond to $\lambda =1/4$. At this point the extra scalar mode is droped out from the phase space (see \cite{bello1} for the 3+1 case). Consequently, the Hamiltonian formulation of this special case is not continuosly connected to the Hamiltonian formulation of the rest of values of $\lambda$. In particular, two additional second-class constraints arise in the Hamiltonian formulation of the kinetic-conformal case. For consistency, any analysis made within the framework of the Hamiltonian formulation must be done separetely for the $\lambda \neq 1/4$ and the $\lambda = 1/4$ cases.

In section \ref{sectiontwo} we present the Ho\v{r}ava-Lifshitz theory in $4+1$ dimensions, we distinguish the $\lambda\neq1/4$ and $\lambda=1/4$ theories. In section \ref{sectionthree} we analyze the $3+1$ theory following a KK reduction. In section \ref{sectionfour} we discuss the gauge vector field equations and compare to the Maxwell equations. In section \ref{sectionfive} we consider in particular the $\lambda=1/4$ theory. Finally in section \ref{sectionsix} we give our conclusions.


\section{Ho\v{r}ava-Lifshitz in five dimensions}
\label{sectiontwo}

We consider a five dimensional manifold foliated by four dimensional Riemannian leaves with metric $G_{\mu\nu} dx^{\mu}\otimes dx^{\nu}$, $\mu,\nu=1,2,3,4$, and parametrized by a time variable. Besides a four vector $N_{\mu}$ and a scalar $N$ under diffeomorphims on the Riemannian foliation are introduced, the shift and the lapse respectively. Both are scalar densities under time reparametrizations.
The Ho\v{r}ava-Lifshitz action on this geometrical framework is given by
\begin{equation}\label{1}
S\left(G_{\mu\nu},N_{\rho},N\right) = \int dtdx^{4} \left[ N \sqrt{G} \bigg( K_{\mu\nu}K^{\mu\nu}-\lambda K^{2}+\beta^{(4)}R+\alpha a_{\mu}a^{\mu} \bigg) 
-V\left(G_{\mu\nu},N\right) \right],   
\end{equation}
where $a_{\mu}=\partial_{\mu}LnN$ and $K_{\mu\nu}K^{\mu\nu}-\lambda K^{2}$ is the kinetic term and the remaining terms describe the potential of the theory. We have written explicitly the lowest order terms in spatial derivatives, the higher order ones are contained in $V\left(G_{\mu\nu},N\right)$, and are constructed from powers of the four dimensional Riemannian tensor and the covariant derivatives of the lapse $N$. The highest order derivative terms of the physical degrees of freedom should be at least of order eight and with an elliptic symbol, in order to obtain a power counting renormalizable theory. In this section we consider the complete theory in $4+1$ dimensions. In the next section, after the KK reduction, in the perturbative analysis we will consider only the dominant terms at large distances, which are the terms of second order in spatial derivatives. This second order truncation in $3+1$ dimensions is related to the Einstein-aether theory \cite{Blas:2009ck,jacobson1,jacobson2,jacobson3}.
$K_{\mu\nu}$ is the extrinsic curvature of the Riemannian leaves,
\begin{equation}\label{2}
K_{\mu\nu}=\frac{1}{2N}\left(\dot{g}_{\mu\nu}-\nabla_{\mu}N_{\nu}-\nabla_{\nu}N_{\mu}\right),    
\end{equation}
and $K$ its trace
\begin{equation}\label{3}
K=G^{\mu\nu}K_{\mu\nu},    
\end{equation} $\lambda$ is a dimensionless coupling constant, while $\alpha$ and $\beta$ are the coupling constants at low energy in the potential of the theory.

We proceed to reformulate the action using the Hamiltonian formalism \cite{kluson,donelly,bello3,bello4}. To do so we have to distinguish the $\lambda\neq1/4 $ and $\lambda=1/4$ cases. At $\lambda=1/4$ the kinetic term is conformal invariant. This is analogous to the $\lambda=1/3$ theory in the Ho\v{r}ava-Lifshitz gravtiy in $3+1$ dimensions which propagates the same degree of freedom as GR and has the same quadrupole radiative behaviour as GR  \cite{bello1,bello2,bello6}. We begin our analysis with the $\lambda \neq 1/4$ case. The conjugate momentum to $G_{\mu\nu}$ is given by
\begin{equation}\label{4}
 \pi^{\mu\nu}=\frac{\partial \mathcal{L}}{\partial \dot{G}_{\mu\nu}}=\sqrt{G}\left(K^{\mu\nu}-\lambda G^{\mu\nu}K\right),   
\end{equation}
and its trace by
\begin{equation}\label{5}
\pi=G_{\mu\nu}\pi^{\mu\nu}=\sqrt{G}\left(1-4\lambda\right)K.    
\end{equation}
Note that if $\lambda = 1/4$, there arise the primary constraint $\pi = 0$. As we commented above, this fact leads to a different Hamiltonian formulation. The Hamiltonian density obtained from the Legendre transformation is 
\begin{equation}\label{6}
\begin{split}
\mathcal{H}=\sqrt{G}N\bigg[\frac{\pi^{\mu\nu}\pi_{\mu\nu}}{G}+\frac{\lambda}{\left(1-4\lambda\right)}\frac{\pi^{2}}{G}-\beta^{(4)}R-\alpha a_{\mu}a^{\mu}\bigg]+2\pi^{\mu\nu}\nabla_{\mu}N_{\nu}+\sigma P_{N} +V\left(G_{\mu\nu},N\right),   
\end{split}
\end{equation}
we have added the $P_N$ term with the Lagrange multiplier $\sigma$ because the theory is subject to a primary constraint given by the conjugate momentum to the lapse $N$ equal to zero, since no time derivatives of $N$ appears in Lagrangian density.

The field equations arising from the canonical Lagrangian are the following. The conservation of the primary constraint, or equivalently variation with respect to $N$ yields the Hamiltonian constraint 
\begin{equation}\label{7}
\frac{\pi^{\mu\nu}\pi_{\mu\nu}}{G}+\frac{\lambda}{\left(1-4\lambda\right)}\frac{\pi^{2}}{G}-\beta^{(4)}R+\alpha a_{\mu}a^{\mu} +2\alpha \nabla_{\mu}a^{\mu}
+\frac{\delta V}{\delta N}
=0, 
\end{equation}
which ends up being a second class constraint.
Variations with respect to $\pi^{\mu\nu}$ gives
\begin{equation}\label{8}
\dot{G}_{\mu\nu}= \frac{2N\pi_{\mu\nu}}{\sqrt{G}}+\frac{2\lambda}{\left(1-4\lambda\right)}G_{\mu\nu}N\frac{\pi}{\sqrt{G}}+\nabla_{\mu}N_{\nu}+\nabla_{\nu}N_{\mu}\,,    
\end{equation}
while variations with respect to $G_{\mu\nu}$ yields
\begin{equation}\label{9}
\begin{split}
-\dot{\pi}^{\mu\nu}=-\frac{1}{2}NG^{\mu\nu}\bigg[\frac{\pi^{\lambda\rho}\pi_{\lambda\rho}}{\sqrt{G}}+\frac{\lambda}{\left(1-4\lambda\right)}\frac{\pi^{2}}{\sqrt{G}}\bigg]+ 2N\bigg[\frac{\pi^{\mu\lambda}\pi^{\nu}_{\lambda}}{\sqrt{G}}+\frac{\lambda}{\left(1-4\lambda\right)}\frac{\pi^{\mu\nu}\pi}{\sqrt{G}}\bigg]   
&\\ +\beta\sqrt{G}N\bigg[^{(4)}R^{\mu\nu}-\frac{1}{2}^{(4)}RG^{\mu\nu}\bigg]-\beta\sqrt{G}\bigg[\nabla^{(\mu}\nabla^{\nu)}N-G^{\mu\nu}\nabla_{\lambda}\nabla^{\lambda}N\bigg]
&\\ -\frac{1}{2}\alpha\sqrt{G}NG^{\mu\nu}a_{\rho}a^{\rho}+\alpha\sqrt{G}Na^{\mu}a^{\nu}+2\nabla_{\rho}\bigg[\pi^{\rho(\mu}N^{\nu)}\bigg]-\nabla_{\rho}\bigg[\pi^{\mu\nu}N^{\rho}
\bigg]
+ \frac{\delta V}{\delta g_{\mu\nu}}
,
\end{split}    
\end{equation}
where 
\begin{equation}
A^{(\mu}B^{\nu)}=\frac{1}{2}\left(A^{\mu}B^{\nu}+A^{\nu}B^{\mu}\right).    
\end{equation}
Finally, variations with respect to $N_{\mu}$ determine the first class constraints of the theory
\begin{equation}\label{10}
\nabla_{\mu}\pi^{\mu\nu}=0.   
\end{equation}
Variations with respect to $P_{N}$ yields $\dot{N}=\sigma$. This equation merely fixes the Lagrange multiplier $\sigma$. 

The degrees of freedom propagated by these non-linear evolution equations can be identified by means of a perturbative analysis. The theory describe six degrees of freedom in $4+1$ dimensions. These are five (transverse-traceless) tensorial modes and one scalar mode. In appendix \ref{app:perturbations} we show a perturbative analysis for the $4+1$ theory.

In the next section we analyze the exact KK reduction of the Ho\v{r}ava-Lifshitz in $4+1$ dimensions. Among other properties we will determine the coupling of Ho\v{r}ava-Lifshitz gravity to the vector gauge potential and two scalar fields.

\section{Non-perturbative Kaluza-Klein reduction to 3+1 dimensions}\label{sectionthree}

In this section we perform the KK reduction of the $4+1$ full Ho\v{r}ava theory in a non-pertubative approach.
We decompose the $4$-dimensional Riemannian metric $G_{\mu\nu}$ in the following form 

\begin{equation}\label{kkdecomposition}
\left(G_{\mu\nu}\right) =\begin{pmatrix}
\gamma_{ij}+\phi A_{i}A_{j} & \quad  \phi A_{j}\\
\phi A_{i} & \phi
\end{pmatrix},
\end{equation}
where $\gamma_{ij}$ is a $3$-dimensional Riemannian metric. We denote $det\left(\gamma_{ij}\right)\equiv\gamma$, thus we have $G\equiv det\left(G_{\mu\nu}\right)=\gamma\phi > 0$, hence $\phi >0$. The inverse metric is then given by 

\begin{equation}\label{inversekkdecomposition}
\left(G^{\mu\nu}\right) =\begin{pmatrix}
\gamma^{ij} &  -A^{j}\\
-A^{i} & \quad \frac{1}{\phi}+A_{k}A^{k}
\end{pmatrix},
\end{equation}
where $\gamma^{ij}$ are the components of the inverse of $\gamma_{ij}$ and $A^{i}=\gamma^{ij}A_{j}$. The decomposition (\ref{kkdecomposition}) is invertible 

\begin{eqnarray}\label{canongammaij}
\gamma_{ij}&=&G_{ij}-\frac{G_{i4}G_{j4}}{G_{44}}\\ \label{canonAi}
A_{j}&=& \frac{G_{4j}}{G_{44}}\\ \label{canonphi}
\phi&=& G_{44}.
\end{eqnarray}

We then have

\begin{equation}\label{legendre}
\pi^{\mu\nu}\dot{G}_{\mu\nu}=\pi^{ij}\dot{\gamma}_{ij}+p^{i}\dot{A}_{i}+p\dot{\phi},    
\end{equation}
where 
\begin{eqnarray}\label{canonpij}
p^{ij}&=&\pi^{ij} \\ \label{canonpi}
p^{i}&=&2 \phi A_{j}\pi^{ij} +2\pi^{i4} \phi\\ \label{canonp}
p&=& \pi^{ij}A_{i}A_{j} +2\pi^{i4}A_{i}+\pi^{44}.
\end{eqnarray}

Equations (\ref{canongammaij})-(\ref{canonp}) define a canonical transformation. In fact,

\begin{equation}\label{poissonbracket}
 \{G_{\mu\nu}\left(x\right),\pi^{\rho\lambda}\left(\tilde{x}\right)\}_{PB} = \frac{1}{2} \left(\delta^{\rho}_{\mu}\delta^{\lambda}_{\nu}+\delta^{\rho}_{\nu}\delta^{\lambda}_{\mu}\right)\delta\left(x-\tilde{x}\right),
\end{equation}
imply 

\begin{equation}\label{poissonbracketgamma}
 \{\gamma_{ij}\left(x\right),p^{kl}\left(\tilde{x}\right)\}_{PB} = \frac{1}{2} \left(\delta^{k}_{i}\delta^{l}_{j}+\delta^{k}_{j}\delta^{l}_{i}\right)\delta\left(x-\tilde{x}\right),
\end{equation}

\begin{equation}\label{poissonbracketAi}
\left\{A_{i}\left(x\right),p^{j}\left(\tilde{x}\right)\right\}_{PB}=\delta^{j}_{i}\delta\left(x-\tilde{x}\right),    
\end{equation}

\begin{equation}
\left\{\phi\left(x\right),p\left(\tilde{x}\right)\right\}_{PB}=\delta\left(x-\tilde{x}\right),    
\end{equation}
and all other Poisson brackets being zero. 

The canonical Lagrangian in $4+1$ can now be reexpressed in terms of the new fields. We then consider the reduced theory by taking $\partial_4 = 0$ on all fields (see appendix \ref{app:higherorderkk}). The reduced canonical Lagrangian is then given by 
\begin{equation}\label{canonlagrangian}
\mathcal{L}=p^{ij}\dot{\gamma}_{ij}+p^{i}\dot{A}_{i}+p\dot{\phi}+P_{N}\dot{N}-\mathcal{H},    
\end{equation}
where the Hamiltonian density is given by
\begin{equation}\label{Hamiltodensity}
\begin{split}
\mathcal{H}= \frac{N}{\sqrt{\gamma\phi}}\bigg[\phi^{2}p^{2}+p^{ij}p_{ij}+\frac{p^{i}p_{i}}{2\phi}+\frac{\lambda}{\left(1-4\lambda\right)}\left(p^{ij}\gamma_{ij}+p \phi\right)^{2}-\gamma\phi\beta^{(4)}R-\gamma\phi\alpha a_{i}a^{i}\bigg] &\\
-\Lambda \partial_{i}p^{i} -\Lambda_{j}\left(\nabla_{i}p^{ij}-\frac{1}{2}p^{i}\gamma^{jk}F_{ik}-\frac{1}{2}p \gamma^{ij}\partial_{i}\phi\right)-\sigma P_{N} +\tilde{V}\left(\gamma_{ij},A_{k},\phi,N\right),
\end{split}
\end{equation}
where $\tilde{V}\left(\gamma_{ij},A_{k},\phi,N\right)$ corresponds to $V\left(G_{\mu\nu},N\right)$ evaluated at $\partial_{4}=0$, that is

\begin{equation}
\tilde{V}\left(\gamma_{ij},A_{k},\phi,N\right)=
V\left(G_{\mu\nu},N\right)|_{\partial_{4}=0}, 
\label{vtilde}   
\end{equation}
and,
\begin{equation}\label{ricci4}
^{(4)}R=R-\frac{\phi}{4}F_{ij}F^{ij}-\frac{2}{\sqrt{\phi}}\nabla_{i}\nabla^{i}\sqrt{\phi},    
\end{equation}
$R$ is the curvature and $\nabla_{i}$ the covariant derivative associated to the $3$-dimensional metric $\gamma_{ij}$. Indices are raised and lowered using $\gamma_{ij}$ and its inverse $\gamma^{ij}$. $\Lambda$ and $\Lambda_{j}$ are the Lagrange multipliers associated to a combination of the constraints (\ref{10}) while $\sigma$ is the Lagrange multiplier associated to the constraint 

\begin{equation}\label{constraintpn}
P_{N}=0    
\end{equation}
that is, the conjugate momentum to $N$ equal zero. These are the primary constraints of the formulation. 

The KK reduction of the momentum constraint (\ref{10}) yields a constraint corresponding to its $4$-component and another one corresponding to the rest spatial directions. After some manipulations, we obtain that these two constraints are equivalent to
\begin{eqnarray}\label{H4}
\mathcal{H}^{4}\equiv\partial_{i}p^{i}&=&0 \\ \label{Hj}
\mathcal{H}^{j}\equiv \nabla_{i}p^{ij}-\frac{1}{2}p^{i}\gamma^{jk}F_{ik}-\frac{1}{2}p\gamma^{ij}\partial_{j}\phi&=&0
\end{eqnarray}
We have then obtained the KK reduction of the full Ho\v{r}ava gravity in $4+1$ dimensions.

We now analize this theory at its $z=1$ limit.
The conservation of these primary constraints 
is satisfied and the conservation of (\ref{constraintpn}) yields the Hamiltonian constraint
\begin{equation}\label{hamiltonianconstraint}
\begin{split}
H_{N}\equiv\frac{1}{\sqrt{\gamma\phi}}\bigg[\phi^{2}p^{2}+p^{ij}p_{ij}+\frac{p^{i}p_{i}}{2\phi}+\frac{\lambda}{\left(1-4\lambda\right)}\left(p^{ij}\gamma_{ij}+p\phi\right)^{2}-\beta\gamma\phi R +\frac{\beta}{4} \gamma \phi^{2} F_{ij}F^{ij} &\\
+2\beta \gamma \sqrt{\phi} \nabla_{i}\nabla^{i} \sqrt{\phi}\bigg]+\alpha\sqrt{\gamma\phi}a_{i}a^{i}+2\alpha\sqrt{\gamma}\nabla_{i}\left(\sqrt{\phi}a^{i}\right)
+\frac{\delta}{\delta N}\left[\tilde{V}\left(\gamma_{ij},A_{k},\phi,N\right)\right]=0.
\end{split}
\end{equation}
The Dirac's procedure to determine the constraints of he theory ends at this step. It turns out that (\ref{H4}) and (\ref{Hj}) are first class constraints while (\ref{constraintpn}) and (\ref{hamiltonianconstraint})
are second class constraints.
The first class constraints, once they are satisfied initially, they are preserved by the evolution equations obtained by taking variations of the action with respect to the independent fields. The second class ones have to be imposed at any time.

We now consider the equations of motion resulting from variations of the canonical action with respect to $p^{ij}$, $p^{i}$ and $p$. They give the equations
\begin{eqnarray}\label{gammadot}
\dot{\gamma}_{ij}&=&\frac{N}{\sqrt{\gamma\phi}}\left[2p_{ij}+\frac{2\gamma_{ij}\lambda}{\left(1-4\lambda\right)}\left(p^{lm}\gamma_{lm}+p\phi\right)\right]+\nabla_{(i}\Lambda_{j)}, \\ \label{Adot}
\dot{A}_{i}&=&\frac{Np_{i}}{\sqrt{\gamma\phi^{3}}}+\partial_{i}\Lambda+\frac{1}{2}\Lambda_{j}\gamma^{jk}F_{ik}\\ \label{phidot}
\dot{\phi}&=&\frac{N}{\sqrt{\gamma\phi}}\left[2p\phi^{2}+\frac{2\lambda}{\left(1-4\lambda\right)}\left(p^{lm}\gamma_{lm}+p\phi\right)\phi\right]+\frac{1}{2}\Lambda^{i}\partial_{i}\phi.
\end{eqnarray}
Variations with respect to $\gamma_{ij}$, $A_{i}$ and $\phi$ yield the equations of motion
\begin{equation}\label{tensorpij}
\begin{split}
\dot{p}^{ij}=\frac{N}{2}\frac{\gamma^{ij}}{\sqrt{\gamma\phi}}\left[\phi^{2}p^{2}+p^{lk}p_{lk}+\frac{1}{\phi}p^{l}p_{l}+\frac{\lambda}{\left(1-4\lambda\right)}\left(p^{lm}\gamma_{lm}+p\phi\right)^{2}\right]  
-\frac{N}{\sqrt{\gamma\phi}}&\\ \times\left[2p^{ik}p^{j}_{k}+\frac{1}{2\phi}p^{i}p^{j}+\frac{2\lambda}{\left(1-4\lambda\right)}\left(p^{lm}\gamma_{lm}+p\phi\right)p^{ij}\right]
+N\sqrt{\gamma\phi}\beta\Bigg[\frac{R}{2}\gamma^{ij}&\\
- R^{ij}\Bigg]+\beta\sqrt{\gamma}\bigg[\nabla^{(i}\nabla^{j)}\left(N\sqrt{\phi}\right)-\gamma^{ij}\nabla_{k}\nabla^{k}\left(N\sqrt{\phi}\right)\bigg]
+\frac{\beta}{2}N\sqrt{\gamma\phi^{3}}&\\ \times\Bigg[F^{in}F^{j}_{n}
-\frac{\gamma^{ij}}{4}F_{mn}F^{mn}\Bigg]
+\beta\sqrt{\gamma}\Bigg[\gamma^{ij}\partial_{l}N\partial^{l}\sqrt{\phi} -2\partial^{i}N \partial^{j}\sqrt{\phi}\Bigg]&\\+\alpha N\sqrt{\gamma\phi} \bigg[\frac{\gamma^{ij}}{2}a_{k}a^{k}
-a^{i}a^{j}\bigg]-\nabla_{k}\bigg[p^{k(i}\Lambda^{j)}-\frac{p^{ij}}{2}\Lambda^{k}\bigg]+\frac{1}{2}\Lambda^{i}p^{l}\gamma^{jm}F_{lm}&\\+\frac{1}{2}p\Lambda^{i}\partial^{j}\phi -\frac{\delta}{\delta\gamma_{ij}}\left[\tilde{V}\left(\gamma_{ij},A_{k},\phi,N\right)\right],
\end{split}    
\end{equation}

\begin{equation}\label{vectorpi}
\dot{p}^{i}=\beta\partial_{j}\left(N\sqrt{\gamma\phi^{3}}F^{ji}\right)-\frac{1}{2}\partial_{k}\left(\Lambda^{k}p^{i}-\Lambda^{i}p^{k}\right)-\frac{\delta}{\delta A_{i}}\left[\tilde{V}\left(\gamma_{ij},A_{k},\phi,N\right)\right],
\end{equation}

\begin{equation}\label{scalarp}
\begin{split}
\dot{p}=-\frac{N}{\sqrt{\gamma}}\Bigg[\frac{3}{2}\sqrt{\phi}p^{2}-\frac{1}{2\sqrt{\phi^{3}}}p^{ij}p_{ij}-\frac{3}{4\sqrt{\phi^{5}}}p^{i}p_{i} +\frac{\lambda}{\left(1-4\lambda\right)}\bigg(\frac{3}{2}\sqrt{\phi}p^{2}&\\+\frac{p p^{ij}\gamma_{ij}}{\sqrt{\phi}}-\frac{1}{2}\frac{\left(p^{ij}\gamma_{ij}\right)^{2}}{\sqrt{\phi^{3}}}\bigg) 
-\gamma\beta\left(\frac{1}{2}\sqrt{\phi}R-\frac{3}{8}\sqrt{\phi}F^{ij}F_{ij}\right)&\\ -\frac{\gamma}{2\sqrt{\phi}}\alpha a_{i}a^{i}\Bigg] -\beta\frac{\sqrt{\gamma}}{\sqrt{\phi}}\nabla_{i}\nabla^{i}N +\frac{1}{2}\partial_{i}\left(p\Lambda^{i}\right)-\frac{\delta}{\delta\phi}\left[\tilde{V}\left(\gamma_{ij},A_{k},\phi,N\right)\right].
\end{split}
\end{equation}

(\ref{H4})-(\ref{scalarp}) are the complete set of field equations of the $3+1$-dimensional theory, it describes the gauge vector-gravity interaction together with two additional scalar fields as already mentioned in the $4+1$-dimension formulation of the previous section.

We may obtain the perturbative equations directly from this $3+1$-formulation (in appendix \ref{app:perturbations} we compare with a perturbative analysis in $4+1$ dimensions). In this analysis we only consider the interacting terms of second order in derivatives. The background is the 3D Euclidean space plus the background conditions for the rest of field variables. They are given by
\begin{equation}
 \hat{\gamma}_{ij} = \delta_{ij} \,,
 \quad
 \hat{p}^{ij} = 0 \,,
 \quad
 \hat{N} = 1 \,,
 \quad
 \hat{A}_i = \hat{p}^i = 0 \,,
 \quad
 \hat{\phi} = 1 \,,
 \quad
 \hat{p} = 0 \,,
 \quad
 \hat{N}_i = \hat{N}_4 = 0 \,.
\end{equation}
 The perturbations around this background are defined by introducing the variables $h_{ij}$, $\Omega_{ij}$, $n$, $n_{i}$ and $n_{4}$ in the following way
\begin{equation}\label{perturbmetric}
\gamma_{ij}=\delta_{ij}+\epsilon h_{ij}, \quad p^{ij}=\epsilon\Omega_{ij}, \quad N_{i}=\epsilon n_{i}, \quad N_{4}=\epsilon n_{4}, \quad N=1+\epsilon n.    
\end{equation}
For the scalar $\phi$ and the vector $A_{i}$ fields we have
\begin{equation}\label{perturbscalarphoton}
A_{i}=\epsilon\xi_{i}, \quad p^{i}=\epsilon\zeta_{i}, \quad \phi=1+\epsilon\tau, \quad p=\epsilon\chi.    
\end{equation}
The perturbative expressions at linear order in $\epsilon$ for the Lagrange multipliers $\Lambda$ and $\Lambda_{i}$ coincide with the corresponding ones to $N_{4}$ and $N_{i}$, respectively. 

The linearized equations of motion become
\begin{eqnarray}\label{scalarphiperturb}
\dot{\tau}&=&2\chi+\frac{2\lambda}{\left(1-4\lambda\right)} \left(\chi +\Omega\right), \\ \label{scalarpperturb}
\dot{\chi}&=&-\frac{\beta}{2}\Delta h-\beta \Delta n,
\end{eqnarray}
\begin{eqnarray}\label{vectorAiperturb}
\dot{\xi}_{i}&=&\zeta_{i}-\partial_{i}n_{4}, \\ \label{vectorpiperturb}
\dot{\zeta}_{i}&=&\beta\partial_{j}\left(\partial_{i}\xi_{j}-\partial_{j}\xi_{i}\right),
\end{eqnarray}
\begin{eqnarray}\label{tensorgammaijperturb}
\dot{h}_{ij}&=&2\Omega_{ij}+\frac{2\delta_{ij}\lambda}{\left(1-4\lambda\right)}\left(\Omega+\xi\right)+2\partial_{(i}n_{j)}, \\ \label{tensorpijperturb}
\dot{\Omega}_{ij}&=&-\frac{\beta}{2}\left(\delta_{ij}-\frac{\partial_{i}\partial_{j}}{\Delta}\right)\Delta h+\frac{\beta}{2}\Delta h_{ij}-\beta\left(\delta_{ij}-\frac{\partial_{(i}\partial_{j)}}{\Delta}\right)\Delta\left(n+\frac{\tau}{2}\right).
\end{eqnarray}
Besides, from the constraints we have
\begin{eqnarray}\label{momenconstraintperturb}
\partial_{i}\Omega_{ij}&=&0\\ \label{hamiltonianconsperturb}
\beta\Delta \tau+2\alpha\Delta n+\beta \Delta h&=&0.
\end{eqnarray}
In order to identify the physical degrees of freedom propagated at linearized level we use the orthogonal transverse/longitudinal decomposition (see Eqs.~(\ref{30}) and (\ref{31}) of  appendix \ref{app:perturbations}), obtaining 
\begin{equation}\label{Aifinal}
\dot{\xi}^{T}_{i}=\zeta^{T}_{i}.    
\end{equation}
\begin{equation}\label{pifinal}
\dot{\zeta}^{T}_{i}=\beta\Delta\xi^{T}_{i},    
\end{equation}
so, combining (\ref{Aifinal}) and (\ref{pifinal}) we get the following wave equation for the vector excitation, 
\begin{equation}\label{photon}
\ddot{\xi}^{T}_{i}-\beta\Delta\xi^{T}_{i}=0. 
\end{equation}
This implies that the vector excitation propagates with speed $\sqrt{\beta}$. From equations (\ref{tensorgammaijperturb}) and (\ref{tensorpijperturb}) we obtain the following wave equation for the graviton
\begin{equation}\label{graviton}
\ddot{h}^{TT}_{ij}-\beta\Delta h^{TT}_{ij}=0 \,.
\end{equation}
We remark that the graviton has the same speed of propagation that the gauge vector, i.e, $\sqrt{\beta}$. The longitudinal modes $\xi^{L}$ and $h^{L}_{i}$ are gauge modes. They are not physical excitations. The remaining 
terms obtained from the decomposition of the equations (\ref{tensorgammaijperturb}) and (\ref{tensorpijperturb}) are
\begin{eqnarray}\label{hpoint}
\dot{h}^{T}&=&2\Omega^{T}+\frac{4\lambda}{\left(1-4\lambda\right)}\left(\Omega^{T}+\chi\right), \\ \label{omegapoint}
\dot{\Omega}^{T}&=&-\frac{\beta}{2}\Delta h^{T}-2\beta\Delta n-\beta\Delta \tau,
\end{eqnarray}
and the longitudinal terms
\begin{equation}\label{longi}
n_{i}+\frac{\lambda}{\left(1-4\lambda\right)}\frac{\partial_{i}}{\Delta}\left(\Omega^{T}+\chi\right)=0,    
\end{equation}
The above equation (\ref{longi}) allows to determine $n_{i}$. So, solving (\ref{hamiltonianconsperturb}) for $\Delta n$ we get
\begin{equation}\label{solvingn}
\Delta n=-\frac{\beta}{2\alpha}\left(\Delta\tau+\Delta h^{T}\right),    
\end{equation}
and combining it with (\ref{scalarphiperturb}), (\ref{scalarpperturb}), (\ref{hpoint}) and (\ref{omegapoint}) we obtain 
\begin{eqnarray}\label{dilaton}
\ddot{h}^{T}-2\ddot{\tau}&=&\beta\Delta\left(h^{T}-2\tau\right)\\ \label{extramode}
\ddot{h}^{T}+\ddot{\tau}&=&\frac{\beta}{\alpha}\frac{\left(1-\lambda\right)\left(3\beta-2\alpha\right)}{\left(1-4\lambda\right)}\Delta\left(h^{T}+\tau\right).  \label{eomextrascalar}  
\end{eqnarray}
Therefore, we have that the $3+1$ theory describes the propagation of six degrees of freedom. They are two transverse-traceless tensorial modes, which are the same modes of GR, two transverse vectorial modes, as in Maxwell theory, and two scalar modes. We may interpret one of these scalar, $h^T + \tau$, as being part of the gravitational interaction of the Ho\v{r}ava theory, that is, the so-called extra mode of the Ho\v{r}ava theory. The other scalar, $h^T - 2\tau$, can be interpreted as part of the coupling to the matter fields resulting from the KK reduction. Indeed, the same KK reduction on GR relativity gives raise to the  $h^T - 2\tau$ scalar in the $3+1$ dimensions. The fact that the factor $1-4\lambda$ arises in the denominator of the right hand side of Eq.~(\ref{eomextrascalar}) signals that the case $\lambda = 1/4$ is a critical point of the theory, as we have commented. The dynamics in this case is different, and not continously connected, to the $\lambda \neq 1/4$ case due to the dropping out of this scalar mode.

\section{The electromagnetic field equations in Ho\v{r}ava formulation}\label{sectionfour}
In this section we analyze in detail the dynamics of the vectorial sector, considering the large distance limit, for which we retain only the $z=1$ terms. This is equivalent to discard the higher order potential $V$. Let us analyze in more details equations (\ref{H4}), (\ref{Adot}) and (\ref{vectorpi}). Since (\ref{H4}) is a first-class constraint it generates a gauge transformation on  $A_{i}$,

\begin{equation}
\delta A_{i}=\{\langle \zeta \partial_{j}p^{j}\rangle,A_{i}\left(x\right)\}_{PB}=
\partial_{i} \zeta\left(x\right),    
\end{equation}
where
\begin{equation}
\langle \zeta \partial_{i}p^{i}\rangle=\int d^{3}\tilde{x} \zeta \left(\tilde{x}\right)\partial_{i}p^{i}\left(\tilde{x}\right).    
\end{equation}
The action and field equations are invariant under this gauge transformation. Note that this holds even including the higher order terms represented by $V$, since the constraint (\ref{H4}) keeps its form unaltered, and it is also independent of all coupling constants. $\Lambda_{i}$ are Lagrange multipliers associated to the first class constraints (\ref{Hj}), they can be fixed to zero in order to simplify the analysis of (\ref{Adot}) and (\ref{vectorpi}). If we denote $A_{0}\equiv\Lambda$ and 
\begin{eqnarray}
 && F_{0i} \equiv \dot{A}_i - \partial_i A_0 \,,
 \\
 && F^{0i} \equiv -\frac{1}{N^{2}}\gamma^{ij}F_{0j} \,,
 \label{foi}     
\end{eqnarray}
then solving $p^{i}$ in (\ref{Adot}) and substituting it in (\ref{vectorpi}), we obtain the equation  
\begin{equation}\label{86}
\partial_{0}\left(F^{0i}N\sqrt{\gamma\phi^{3}}\right)+\beta\partial_{j}\left(F^{ji}N\sqrt{\gamma\phi^{3}}\right)=0.
\end{equation}

We now may compare with the standard Maxwell theory written in relativistic variables. To this end, using the field variables of the $3+1$ Ho\v{r}ava theory, we may build a four-dimensional metric $g_{\mu\nu}$ decomposed in the standard ADM way,
\begin{eqnarray}\label{ADM} \nonumber
g_{00}&=&-N^{2}+\gamma_{ij}N^{i}N^{j} \\
g_{0i}&=&g_{i0}=N_{i}, \quad N_{i}=\gamma_{ij}N^{j} \\ \nonumber
g_{ij}&=&\gamma_{ij},
\end{eqnarray}
with inverse given by 
\begin{eqnarray}\label{inverseADM} \nonumber
g^{00}&=&-\frac{1}{N^{2}}\\
g^{0i}&=&g^{i0}=\frac{N^{i}}{N^{2}}\\ \nonumber
g^{ij}&=&\gamma^{ij}-\frac{N^{i}N^{j}}{N^{2}} \,.
\end{eqnarray}
For simplicity, if we fix $N_{i}=0$ as a gauge of the space-like diffeomorphisms on the side of the relativistic theory, then we have that from our definition of $F^{0i}$ (\ref{foi}),  
\begin{equation}
F^{0i}=g^{0\mu}g^{i\nu}F_{\mu\nu},    
\end{equation}
$F_{\mu\nu}$ being the four-dimensional definition of the curvature of the potential $A_{\mu}$, where $A_{0}$ is the Lagrange multiplier of the constraint (\ref{H4}), the generator of gauge transformation.

The non-relativistic electromagnetic equations are exactly the relativistic ones if $\beta=1$. In fact, $N\sqrt{\gamma}=\sqrt{|g|}$ and (\ref{86}) becomes the relativistic equations of the electromagnetic field coupled to the dilaton field. If in addition $\phi=1$, we have that Eq.~(\ref{86}) becomes
\begin{equation}\label{90}
\partial_{\mu}\left(\sqrt{|g|}F^{\mu i}\right)=0,    
\end{equation}
while (\ref{H4}) can be expressed, when $\phi=1$, as 
\begin{equation}\label{91}
\partial_{\mu}\left(\sqrt{|g|}F^{\mu 0}\right)=0.
\end{equation}
The speed of propagation of the gauge vector excitation is in general $\sqrt{\beta}$ and it is the same for the gravity excitation. It is interesting that this property is a consequence of starting with a purely gravitational FDiff-invariant theory in $5$-dimensions. We can now identify the gauge vector with the electromagnetic potential. Equations (\ref{86}) and (\ref{91}) are the anisotropic electromagnetic equations. We remark again that (\ref{91}), arising directly from (\ref{H4}), does not involve any coupling constant. Since the potential in the Hamiltonian of the $4+1$ complete theory is constructed from polynomial expressions in terms of the Riemann tensor  and the lapse function and its covariant derivatives up to $2z$ derivatives, its reduction to $3+1$ dimensions is a polynomial (of finite degree) in the gauge vector and its covariant derivatives.

\section{The kinetic conformal point in 4+1 dimensions}\label{sectionfive}

We discuss in this section the Ho\v{r}ava-Lifshitz gravity in $4+1$ dimensions for $\lambda=1/4$ and its $3+1$ reduction. It is a different theory with respect to the $\lambda\neq1/4$ formulation we have already considered. The  propagating degrees of freedom are different in the two cases. The corresponding relation in Ho\v{r}ava-Lifshitz gravity in $3+1$ dimensions is between the $\lambda=1/3$ and $\lambda\neq1/3$ theories \cite{bello2,bello1}. In the canonical analysis in section \ref{sectiontwo}, in the case $\lambda=1/4$, we obtain an additional primary constraint 
\begin{equation}\label{momentumconstraint}
\pi=G_{\mu\nu}\pi^{\mu\nu}=0, \quad \mu,\nu=1,2,3,4.    
\end{equation}
The Hamiltonian density in this case is given by 
\begin{equation}\label{hamiltondensity}
\mathcal{H}=N\sqrt{G}\left[\frac{\pi^{\mu\nu}\pi_{\mu\nu}}{G}-\beta^{(4)}R-\alpha a_{\mu}a^{\mu}\right]+2\pi^{\mu\nu}\nabla_{\mu}N_{\nu} +\sigma P_{N} +\mu\pi
+ V,
\end{equation}
where $\sigma$ and $\mu$ are Lagrange multipliers associated to two primary constraints, as before $P_{N}$ is the conjugate momentum to $N$. Additionally 
\begin{equation}\label{firstclass}
\nabla_{\mu}\pi^{\mu\nu}=0    
\end{equation}
is the primary constraint associated to the invariance under the diffeomorphisms on the Riemannian leaves of the $4+1$ foliation. It is a first-class constraint, it is preserved under the evolution determined by the Hamiltonian.

The conservation of the primary constraints (\ref{momentumconstraint}) and $P_{N}=0$ implies two new constraints, the Hamiltonian constraint
\begin{equation}\label{hamiltocons}
\frac{\pi^{\mu\nu}\pi_{\mu\nu}}{G}-\beta^{(4)}R+\alpha a_{\mu}a^{\mu}+2\alpha \nabla_{\mu}a^{\mu}
+\frac{\delta V}{\delta N}
=0,    
\end{equation}
and
\begin{equation}\label{piconservation}
2\frac{\pi^{\mu\nu}\pi_{\mu\nu}}{G}+\beta^{(4)}R+\left(\alpha-3\beta\right)a_{\mu}a^{\mu}-3\beta\nabla_{\mu}a^{\mu}
- G_{\mu\nu} \frac{\delta V}{\delta G_{\mu\nu}}
=0.    
\end{equation}
The conservation of (\ref{hamiltocons}) and (\ref{piconservation}) determine Lagrange multipliers in the formulation. The Dirac's procedure to determine the complete set of constraint of the theory ends at this stage: (\ref{momentumconstraint}), $P_{N}=0$, (\ref{hamiltocons}) and (\ref{piconservation}) are second class constraints, (\ref{firstclass}) are first class constraints.

We now consider the complete set of field equations of the theory. In order to derive them, we assume an asymptotic decay to zero of the Lagrange multipliers of the second class constraints. It is then correct to use the Hamiltonian (\ref{hamiltondensity}) without the explicit introduction of the whole set of second class constraints via Lagrange multipliers \cite{bello5}. The field equations are, together with (\ref{momentumconstraint}), (\ref{firstclass}), (\ref{hamiltocons}) and (\ref{piconservation}), the following ones,
\begin{equation}\label{107}
\dot{G}_{\mu\nu}=\frac{2N\pi_{\mu\nu}}{\sqrt{G}}+ \mu G_{\mu\nu}+\nabla_{\mu}N_{\nu}+\nabla_{\nu}N_{\mu},    
\end{equation}

\begin{equation}\label{108}
\begin{split}
-\dot{\pi}^{\mu\nu}=-\frac{1}{2}NG^{\mu\nu}\frac{\pi^{\lambda\rho}\pi_{\lambda\rho}}{\sqrt{G}}+ 2N\frac{\pi^{\mu\lambda}\pi^{\nu}_{\lambda}}{\sqrt{G}}  
 +\beta\sqrt{G}N\bigg[^{(4)}R^{\mu\nu}-\frac{1}{2}^{(4)}RG^{\mu\nu}\bigg]&\\-\beta\sqrt{G}\bigg[\nabla^{(\mu}\nabla^{\nu)}N-G^{\mu\nu}\nabla_{\lambda}\nabla^{\lambda}N\bigg]
 -\frac{1}{2}\alpha\sqrt{G}NG^{\mu\nu}a_{\rho}a^{\rho}+\alpha\sqrt{G} &\\ 
 \times N a^{\mu}a^{\nu}
 +2\nabla_{\rho}\bigg[\pi^{\rho(\mu}N^{\nu)}\bigg]-\nabla_{\rho}\bigg[\pi^{\mu\nu}N^{\rho}\bigg] +\mu\pi^{\mu\nu}
 + \frac{\delta V}{\delta G_{\mu\nu}} \,.
\end{split}    
\end{equation}

We now introduce the KK ansatz given by (\ref{kkdecomposition}) and (\ref{inversekkdecomposition}). We can perform the canonical transformation defined in section \ref{sectionthree} and impose $\partial_{4}=0$ on the fields to obtain a $3+1$ formulation. The Hamiltonian in this formulation becomes 

\begin{equation}\label{Hamiltodensitylambdafixed}
\begin{split}
\mathcal{H}= \frac{N}{\sqrt{G}}\bigg[\phi^{2}p^{2}+p^{ij}p_{ij}+\frac{p^{i}p_{i}}{2\phi}-\beta G ^{(4)}R-\alpha G a_{\mu}a^{\mu}\bigg] 
-2\bigg[\nabla_{\mu}\pi^{\mu\nu}\bigg]N_{\nu} &\\ +\sigma P_{N}+\mu\bigg[p^{ij}\gamma_{ij}+p\phi\bigg]
+ \tilde{V} \,,
\end{split}
\end{equation}
where $^{(4)}R$ has the expression (\ref{ricci4}) and $\tilde{V}$ is defined in (\ref{vtilde}).

A perturbative analysis can be done in order to identify the physical degrees of freedom. In this case we do not consider the higher order terms. Following the analysis in section \ref{sectionthree}, we obtain the perturbative equations
\begin{eqnarray}\label{103}
\ddot{h}^{TT}_{ij}&=&\beta \Delta h^{TT}_{ij} \\ \label{104}
\ddot{A}^{T}_{i}&=&\beta \Delta A^{T}_{i} \\ \label{105}
\ddot{\phi}&=&\beta \Delta \phi\\ \label{106}
h^{T}&=&\phi,
\end{eqnarray}
that is, comparing with the $\lambda\neq1/4$ theory there is only one propagating scalar field. All physical modes propagate with the same speed $\sqrt{\beta}$. The longitudinal components $h^{L}_{i}$, $A^{L}$ are gauge modes. We notice that the propagating degrees of freedom are the same to the ones in GR in interaction with the electromagnetic and the dilaton fields arising from a KK reduction of GR in $4+1$ dimensions. Although the Ho\v{r}ava's theory at the kinetic conformal point we are considering breaks the relativistic symmetry, it propagates the same degrees of freedom as the corresponding one in GR. 

The non-relativistic electromagnetic field equations are the same as in the previous section, however the other field equations are different, in particular a new second class constraint appears in this formulation compared to the $\lambda\neq1/4$ theory. 

It is important to discuss the field equations obtained from the Hamiltonian in this section, equation (\ref{Hamiltodensitylambdafixed}), and in section \ref{sectionthree} in comparison to the field equations arising from an action in which it is assumed that the scalar field $\phi$ is in its ground state, which we take to be $\phi=1$, $p=0$. In fact, variations of an action subject to the restriction $\phi=1$, $p=0$ give rise to field equations which are not equivalent to the ones in this section or in section \ref{sectionthree} on which one imposes the $\phi=1$, $p=0$ restriction. 

It is straightforward to obtain the equations from the action restricted
by $\phi=1$ and $p=0$. In fact, variations with respect to $\gamma_{ij}$, $p^{ij}$, $A_{i}$, $p^{i}$ determine the same field equations obtained by taking variations of the canonical Lagrangian associated to (\ref{Hamiltodensitylambdafixed}), in the $\lambda=1/4$ case, and imposing afterwards $\phi=1$, $p=0$. The main difference being that in the restricted case there are not field equations corresponding to variations on $\phi$ and its conjugate momenta. The analysis of the field equations in the restricted case show that the only physical degrees of freedom in the theory correspond to the $h^{TT}_{ij}$ tensorial modes and the $A^{T}_{i}$ vectorial modes. The corresponding perturbation equations are (\ref{103}) and (\ref{104}).

\section{Conclusions}\label{sectionsix}

We have obtained the gravitational theory of Ho\v{r}ava in $3+1$ dimensions coupled to vector and scalar fields in a FDiff-invariant way. We have used a procedure of dimensional reduction starting with a purely gravitational nonprojectable Ho\v{r}ava theory in $4+1$ dimensions. These are in principle power counting renormalizable theories provided all $z=4$ interaction terms are included in the potential. This a relevant point because in some cases the coupling to matter fields may damage the behaviour of the divergences of the vacuum theory. For example it is known \cite{tHooft} that pure GR is finite at one loop but it is badly divergent at one loop when a scalar field is coupled to it. 

In the $3+1$ coupled theory there arise a gauge symmetry in the vectorial sector, equivalent to the gauge symmetry of the Maxwell electromagnetism on a curved background. The gauge symmetry is generated by the same first class constraint as in the electromagnetic-gravity theory in GR. Moreover, the field equations for the gauge vector have the same structure as in the relativistic case. In particular if we take $\beta=1$, the Ho\v{r}ava-Lifshitz theory is still an anisotropic formulation of the gravity-electromagnetic interaction in which the field equations for the gauge vector are exactly the Maxwell equations on a curved background. By means of a perturbative approach, we have identified the excitations of the gravitational, vector and scalar fields. With respect to the speed of propagation of the graviton and the gauge vector field, it is the same velocity $\sqrt{\beta}$ for both excitations.

We have obtained the Hamiltonian, field equations of the theory and determine the propagating degrees of freedom. We have separated the $\lambda \neq 1/4$ and the $\lambda = 1/4$ cases, since they correspond to two different dynamics of the physical degrees of freedom. In the $\lambda \neq 1/4$ case, the physical degrees of freedom are the same as the ones in the gravity+electromagnetic+dilaton interaction described in GR, plus an additional scalar mode. In the $\lambda=1/4$ theory, the additional scalar mode is absent, because the different structure of the constraints. This is an extension of same result in \cite{bello1,bello2} for the pure Ho\v{r}ava-Lifshitz gravity at the kinetic conformal point, $\lambda=1/3$ in that case. It is interesting that the introduction of the interaction of gravity with the gauge vector and dilaton fields shifts the kinetic conformal point from $\lambda=1/3$ to $\lambda=1/4$. This is an important point. 

A further step could be to analyze explicitly the higher order terms in the potential and to analyze the non-linear, non-relativistic extension of the Maxwell equations. We remark that the new non-linear interaction terms are necessarily polynomial of \emph{finite} degree in the gauge vector field and its covariant derivatives, since the metric and its inverse are polynomial in the gauge vector field. This is different to the couplings arising in some other theories, like the Born-Infeld theory, where the expansion in the field strenght is an infinite serie.

We have also commented on the case when the dilaton field is in its ground state. That is, the case when $\phi$ is constant in the Lagrangian. The theory in that case propagates only the gravity and electromagnetic excitations, the transverse traceless components of the metric and the transverse gauge vector components just as in GR or Maxwell equations.


\section*{Acknowledgements}
A. Restuccia is partially supported by grant Fondecyt No. $1161192$, Chile. F. Tello-Ortiz thanks the financial support of the project ANT1756 and the Phd program at the Universidad de Antofagasta.


\appendix

\section{Kaluza-Klein reduction for Lifshitz-like theories}
\label{app:higherorderkk}
We take an anisotropic scalar-field model represented by the real scalar field $\phi$, which depends on $D+1$ spatial dimensions and the time $t$. The spatial base manifold is a $D+1$ flat manifold that has a compact component of large $2\pi L$. We use the coordinate $y$ to parameterize the compact dimension, whereas $x^i$ labels the rest of spatial coordinates. We consider a Lifshitz model of $z=2$ order of anisotropy. The action of the model is
\begin{equation}
	S = \int dt d^Dx \int\limits_0^{2 \pi L} dy
	\left(
	( \partial_t \phi )^2 + \alpha \phi \tilde{\Delta} \phi
	+ \beta \phi \tilde{\Delta}^2 \phi 
	\right)  \,,
\label{lifshitzmodel}
\end{equation}
where $\tilde{\Delta}$ is the flat spatial Laplacian in the $D+1$ spatial dimensions,
\begin{equation}
 \tilde{\Delta} = \Delta + \partial_{yy} \,,
\end{equation}
$\Delta$ is the flat Laplacian in the $D$ dimensions, and $\alpha,\beta$ are coupling constants.

We introduce the Fourier expansion of the scalar field, namely
\begin{equation}
 \phi(t,x^i,y) = 
 \sum\limits_{m=-\infty}^{\infty} \phi_m(t,x^i) e^{\frac{i m}{L} y} \,,
\end{equation}
where the modes $\phi_m$ are restricted in order to the field $\phi(t,x^i,y)$ be real. The action for the Fourier modes becomes
\begin{equation}
\begin{array}{rcl}
 S &=& {\displaystyle
 2\pi L \sum\limits_{m=-\infty}^{\infty} \int dt d^Dx \left[
  - \phi_m \ddot{\phi}_m 
  + \alpha \phi_m \left( \Delta - \left(\frac{m}{L}\right)^2 \right) \phi_m
  + \beta \left( \phi_m \Delta^2 \phi_m \right.\right.
} \\[1ex]  && {\displaystyle
   \left.\left.
      - 2 \left(\frac{m}{L}\right)^2 \phi_m \Delta \phi_m
      + \left(\frac{m}{L}\right)^4 \phi_m^2 \right)\right]
}
\end{array} \,.
\end{equation}
We emphasize that this is equal to the action (\ref{lifshitzmodel}) since the Fourier expansion for $\phi \in C^{\infty}(\Sigma)$ converges pointwise. The field equation for $m\neq 0$ is given by
\begin{equation}
 - \ddot{\phi}_m + \alpha \Delta \phi_m + \beta \Delta^2 \phi_m
 - \alpha \left(\frac{m}{L}\right)^2 \phi_m
 - 2\beta \left(\frac{m}{L}\right)^2 \Delta \phi_m
 + \left(\frac{m}{L}\right)^4 \phi_m = 0 \,. 
\end{equation}
For each given $m \neq 0$, we take the $L \rightarrow 0$ limit. In this equation there are several coefficients that diverge, $\left(\frac{m}{L}\right)^4$ being the one of highest order. Therefore, for the existence of the solution in this limit it is neccessary that $\phi_m = 0$ for all $m \neq 0$. 

The field equation for $m = 0$ is
\begin{equation}
  - \ddot{\phi}_0 + \alpha \Delta \phi_0 + \beta \Delta^2 \phi_0 = 0 \,.
\end{equation}
This equation can be obtained from the following action in one dimension less,
\begin{equation}
 S_{\mbox{\tiny KK}} =
 \int dt d^Dx \left(
 - \phi_0 \ddot{\phi}_0 + \alpha \phi_0 \Delta \phi_0
 + \beta \phi_0 \Delta^2 \phi_0 
 \right) \,.
\end{equation}


\section{The degrees of freedom in $4+1$ dimensions}
\label{app:perturbations}
Here we perform a linear-order perturbative analysis on the $4+1$ theory around an Euclidean four dimensional background together with the backgorund conditions for the rest of canonical variables. In this analysis we only consider the $z=1$ terms, hence we neglect the interacting terms in $V$. The background is given by 
\begin{equation}\label{11}
G_{\mu\nu}=\delta_{\mu\nu}, \quad \pi^{\mu\nu}=0, \quad \mbox{together with}\quad N_{\mu}=0, \quad N=1 \,.
\end{equation}
We introduce the pertubative variables according to
\begin{eqnarray}\nonumber
G_{\mu\nu}&=&\delta_{\mu\nu}+k_{\mu\nu} \\ \label{12}
\pi^{\mu\nu}&=&\Omega^{\mu\nu} \\ \nonumber
N_{\mu}&=&n_{\mu} \\ \nonumber
N&=&1+n \,.
\end{eqnarray}
We obtain to first order 
\begin{eqnarray}\label{13}
^{(4)}R_{\mu\nu}&=&\frac{1}{2}\partial_{\lambda}\partial_{\nu}k_{\mu\lambda}+\frac{1}{2}\partial_{\mu}\partial_{\lambda}k_{\nu\lambda}-\frac{1}{2}\partial_{\mu}\partial_{\nu}k_{\lambda\lambda}-\frac{1}{2}\Delta k_{\mu\nu} \\ \label{14}
^{(4)}R&=&\partial_{\mu}\partial_{\lambda}k_{\mu\lambda}-\Delta k_{\mu\mu}.
\end{eqnarray}
We will use the $T+L$ decomposition of the four dimensional tensors, 
\begin{equation}\label{15}
k_{\mu\nu}=k^{TT}_{\mu\nu}+\frac{1}{3}\left(\delta_{\mu\nu}-\frac{\partial_{\mu}\partial_{\nu}}{\Delta}\right) k^{T}+\partial_{\nu}k^{L}_{\mu}+\partial_{\mu}k^{L}_{\nu}.   
\end{equation}
We obtain,
\begin{equation}\label{16}
^{(4)}R_{\mu\nu}-\frac{1}{2}G_{\mu\nu}^{(4)}R=-\frac{1}{2}\Delta k^{TT}_{\mu\nu}+\frac{1}{3}\left(\delta_{\mu\nu}-\frac{\partial_{\mu}\partial_{\nu}}{\Delta}\right)\Delta k^{T}.    
\end{equation}
From (\ref{10}) we obtain
\begin{eqnarray}\label{17}
\Omega^{L}&=&0\\ \nonumber
\Omega_{\mu\nu}&=&\Omega^{TT}_{\mu\nu}+\frac{1}{3}\left(\delta_{\mu\nu}-\frac{\partial_{\mu}\partial_{\nu}}{\Delta}\right)\Omega^{T}.
\end{eqnarray}
From (\ref{8}) we get
\begin{eqnarray}\label{18}
\dot{k}^{TT}_{\mu\nu}&=&2\Omega^{TT}_{\mu\nu}\\ \label{19}
\dot{k}^{T}&=&\frac{2\left(1-\lambda\right)}{\left(1-4\lambda\right)}\Omega^{T}\\ \label{20}
\dot{k}^{L}_{\mu}&=&N_{\mu}+\frac{\lambda}{\left(1-4\lambda\right)}\frac{\partial_{\mu}}{\Delta}\Omega^{T} \,.
\end{eqnarray}
From (\ref{9}) we get
\begin{eqnarray}\label{21}
\dot{\Omega}^{TT}_{\mu\nu}&=&\frac{\beta}{2}\Delta k^{TT}_{\mu\nu} \\ \label{22}
\dot{\Omega}^{T}&=&-\beta \Delta k^{T}-3\beta\Delta n,
\end{eqnarray}
and from (\ref{7})
\begin{equation}\label{23}
\beta \Delta k^{T}+2\alpha\Delta n=0.  
\end{equation}
Finally, we may combine the above equations to obtain 
\begin{eqnarray}\label{24}
\ddot{k}^{TT}_{\mu\nu}&=&\beta\Delta k^{TT}_{\mu\nu} \\ \label{25}
\ddot{k}^{T}&=&\beta \frac{\left(1-\lambda\right)}{\left(1-4\lambda\right)}\frac{\left(3\beta-2\alpha\right)}{\alpha}\Delta k^{T},
\end{eqnarray}
we remark that these equations are gauge independent. They describe the propagation of the six degrees of freedom of the theory. In particular
(\ref{24}) describes the evolution of $5$ tensorial degrees of freedom while (\ref{25}) describes the propagation of a scalar one. It follows that when reducing \emph{a la} KK these field equations to a $3+1$ dimensions, the $5$ degrees of freedom decompose into $2+2+1$ corresponding to the graviton, two degrees of freedom of a vector gauge field and one degree of freedom of a scalar field, all of them propagating with the same velocity $\sqrt{\beta}$. In order to show this decomposition we assume the fields are independent of one spatial coordinate. We first invert the $T+L$ decomposition, to obtain
\begin{eqnarray}\label{26}
k^{L}_{\nu}&=&\frac{1}{\Delta}\partial_{\rho}k_{\rho\nu}-\frac{1}{2}\frac{\partial_{\nu}}{\Delta}\left(\frac{\partial_{\lambda}\partial_{\rho}k_{\lambda\rho}}{\Delta}\right) \\ \label{27}
k^{T}&=&k_{\mu\mu}-\frac{\partial_{\rho}\partial_{\lambda}h_{\rho\lambda}}{\Delta}\\ \label{28}
k^{TT}_{\mu\nu}&=&k_{\mu\nu}-\frac{1}{3}\left(\delta_{\mu\nu}-\frac{\partial_{\mu}\partial_{\nu}}{\Delta}\right)k^{T}-\partial_{\mu}k^{L}_{\nu}-\partial_{\nu}k^{L}_{\mu}.
\end{eqnarray}
We now decompose $k_{\mu\nu}$ into
\begin{eqnarray}\nonumber
k_{ij}&=&h_{ij} \\ \label{29}
k_{4i}&=&A_{i}\\ \nonumber
k_{44}&=&\phi,
\end{eqnarray}
and perform a three dimensional $T+L$ decomposition of $h_{ij}$ and $A_{i}$
\begin{eqnarray}\label{30}
h_{ij}&=&h^{TT}_{ij}+\frac{1}{2}\left(\delta_{ij}-\frac{\partial_{i}\partial_{j}}{\Delta}\right)h^{T}+\partial_{i}h^{L}_{j}-\partial_{j}h^{L}_{i} \\ \label{31}
A_{i}&=&A^{T}_{i}+\partial_{i}A^{L}.
\end{eqnarray}
By replacing (\ref{30}) and (\ref{31}) into (\ref{26}), (\ref{27}) and (\ref{28}), we obtain 
\begin{eqnarray}\nonumber
k^{L}_{i}&=&h^{L}_{i} \\ \nonumber
k^{L}_{4}&=&A^{L} \\ \label{32}
k^{T}&=&h^{T}+\phi \\ \nonumber
k^{TT}_{4i}&=& A^{T}_i \\ \nonumber
k^{TT}_{44}&=&\frac{1}{3}\left(2\phi-h^{T}\right) \\ \nonumber
k^{TT}_{ij}&=&h^{TT}_{ij}-\frac{1}{6}\left(\delta_{ij}-\frac{\partial_{i}\partial_{j}}{\Delta}\right)\left(2\phi-h^{T}\right).
\end{eqnarray}
Finally, from (\ref{24}) we have 
\begin{eqnarray}\label{33}
\ddot{h}^{TT}_{ij}&=&\beta\Delta h^{TT}_{ij} \\ \label{34}
\ddot{A}^{T}_i &=& \beta \Delta A^{T}_i \\ \label{35}
2\ddot{\phi}-\ddot{h}^{T}&=&\beta\Delta\left(2\phi-h^{T}\right).
\end{eqnarray}
(\ref{33}) describes the propagation of the two degrees of freedom of the graviton, (\ref{34}) of the two degrees of freedom of the gauge vector and (\ref{35}) of the scalar field $2\phi-h^T$, all of them propagating with velocity $\sqrt{\beta}$. In addition, we have one propagating scalar degree of freedom described by (\ref{25}). This is in complete agreement with the results in section III, where the perturbative analysis was done in the $3+1$ theory.


\end{document}